# Interplay of Probabilistic Shaping and Carrier Phase Recovery for Nonlinearity Mitigation

Stella Civelli [(1,2)], Enrico Forestieri[(1,2)], Marco Secondini[(1,2)]

[(1)] Tecip Institute, Scuola Superiore Sant'anna stella.civelli@santannapisa.it
[(2)] PNTLab, Consorzio nazionale interuniversitario per le telecomunicazioni (CNIT)

***Abstract*** *The interaction between carrier phase recovery and probabilistic amplitude shaping (PAS) in the nonlinear regime is investigated. We show that, for sufficiently high signal-to-noise ratio, the first provides the same inter-channel nonlinearity mitigation achieved by short block-length PAS.*

## Introduction

Probabilistic amplitude shaping (PAS) has been recently widely investigated to improve the performance of the optical fiber network. PAS allows to finely adapt the information rate to the system requirements (channel signal-to-noise ratio (SNR) and forward error correction (FEC) code) and to reduce the gap to the Shannon limit in the linear regime[1],[2]. The SNR gain—up to $1.53$ dB for large constellation size[3]—depends on the particular implementation of PAS, handled by the distribution matcher (DM). The DM maps $k$ independent input bits with uniform distribution to $N$ *correlated* output amplitudes with the desired Maxwell–Boltzmann (MB) distribution. While different implementations of DM exist, its performance generally improves with the block length $N$. In fact, as $N$ increases, the correlation between the symbols of each block decreases, allowing to encode more information on each transmitted symbol. For $N \to \infty$, the DM output looks like an i.i.d. source with MB distribution, yielding the optimal PAS gain in the linear regime for a given rate and constellation size[3].

While most of the recent studies on PAS concerned the DM implementation and PAS performance in the linear regime[2],[4]–[8], some studies on the impact of PAS in the nonlinear regime have been done[1],[9],[10]. In particular, it has been shown that the correlation induced by PAS on the output symbols reduces the amplitude fluctuations inside each block, reducing the amount of nonlinear interference generated by each channel and yielding an additional *nonlinear shaping gain*. In this case, the amplitude correlation induced by the DM is beneficial, so that the nonlinear shaping gain decreases as $N$ increases, and vanishes for $N \to \infty$. Therefore, there is an optimal block length which maximizes the shaping gain by providing the best trade-off between linear and nonlinear gain.

So far, this effect has been mainly investigated in the absence of any carrier phase recovery algorithm. However, it is known that a good part of the inter-channel nonlinear interference generated by amplitude fluctuations manifests as correlated phase noise, which can be alternatively mitigated by a properly optimized carrier phase recovery algorithm[11]–[15] (and references therein). Thus, the question is whether, in the presence of carrier phase recovery—always required in practical systems due to the phase noise of the transmission and detection lasers—PAS still has an optimal block length and provides an additional nonlinear shaping gain. In this work, we investigate by simulations the interaction between carrier phase recovery and PAS in a wavelength-division multiplexing (WDM) scenario, by considering a common blind phase search (BPS) algorithm[16] for carrier phase recovery and implementing PAS through the enumerative sphere shaping (ESS) algorithm[5].

## System setup and results

The system setup is sketched in Fig. 1. A stream of uniformly distributed bits—representing the information bits after FEC encoding—feeds the PAS block, which maps the bits to symbols of a $256$ quadrature amplitude modulated (QAM) constellation with a desired rate. In particular, for each block of $k$ bits, the DM generates $N$ amplitudes that, together with $N$ uniformly distributed bits, generate the in-phase and quadrature components of $N/4$ consecutive $2$-polarization $256$-QAM symbols. Using a root raised cosine (RRC) pulse with rolloff $0.1$ and baud rate $R_s = 41.67$ GBd, the signals corresponding to $4$ adjacent channels are multiplexed in a single superchannel, the superchannel of interest (SCOI), with $75$ GHz spacing. Two additional superchannels, with the same properties of the SCOI, are also multiplexed, such that 12 evenly spaced channels are transmitted over an overall bandwidth of $900$ GHz. The generated

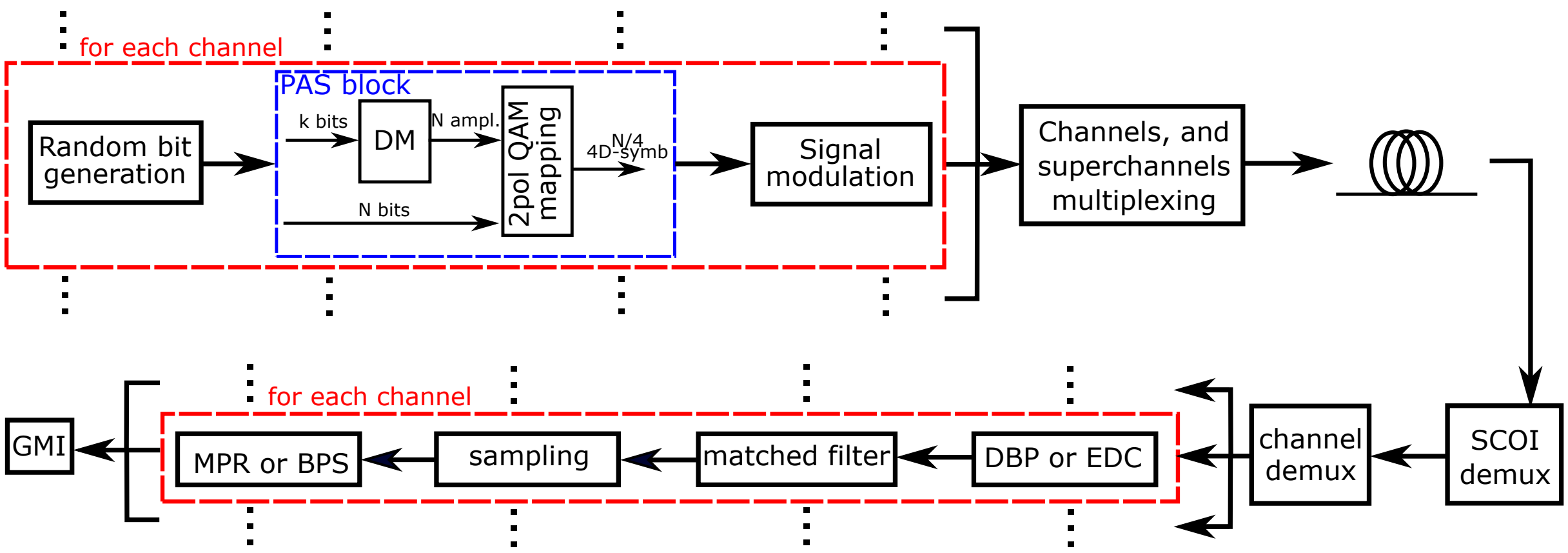

**Fig. 1:** Simulation setup

WDM waveform is launched into the link, composed of several spans of $80$ km single mode fiber (dispersion $D = 17$ ps/nm/km, Kerr parameter $\gamma = 1.3\,\mathrm{W}^{-1}\mathrm{km}^{-1}$, attenuation $\alpha_{\mathrm{dB}} = 0.2$ dB/km). After each span, an erbium-doped fiber amplifier (EDFA) with noise figure $5$ dB compensates for loss. At the end of the link, the side superchannels are filtered out, and the $4$ channels of the SCOI are demultiplexed. Each channel undergoes: (i) either ideal digital back propagation (DBP) or electronic dispersion compensation (EDC), (ii) matched filtering, (iii) sampling at symbol time, and (iv) nonlinear phase noise mitigation. Finally, the average generalized mutual information (GMI) of the $4$ channels of the SCOI is evaluated, representing the average information per symbol that can be reliably transmitted on each polarization and channel of the SCOI, assuming an ideal FEC and bit-wise decoding[6],[17].

As regards the nonlinear phase noise mitigation, two different approaches are considered: blind phase search (BPS) and mean phase rotation (MPR). On the one hand, BPS is a practical carrier phase recovery algorithm typically employed to mitigate laser phase noise for QAM constellations[16]. In a nutshell, each symbol is rotated by an angle which minimizes the mean square error between the rotated symbol and its decision on $2N_{\mathrm{BPS}} + 1$ consecutive symbols. We consider an optimized $N_{\mathrm{BPS}}$ and 64 test angles in a $\pi/2$ interval, though the latter can be reduced to alleviate the computational cost. On the other hand, MPR consists in simply rotating all the symbols by the same average phase (one for each polarization), in practice corresponding to a BPS with $N_{\mathrm{BPS}} \to \infty$. This is the typical approach employed in simulations—when the laser phase noise is not considered and, thus, carrier phase recovery is not required—to simply remove the average phase rotation induced by fiber nonlinearity for a given total launch power.

The PAS block implements sphere shaping, consisting in mapping bit sequences to amplitude sequences with minimum energy, by the enumerative sphere shaping (ESS) algorithm[5],[18],[19]. For a given block length $N$ and constellation size, sphere shaping maximizes the source rate for a given average power, yielding the best performance in the linear regime. For $N \to \infty$, the PAS output approaches an i.i.d. source with MB distribution[3], which yields the ultimate linear shaping gain. In this work, the PAS rate is $6$ bits/symbol—equivalent to a DM rate of $2$ bits/amplitude.

Fig. 2 shows the maximum average GMI (for an optimized launch power) versus shaping block length $N$, with or without ideal DBP, and with or without BPS, when $15$ spans of fiber are considered. The corresponding total achievable information rate for the SCOI (two polarizations and four channels) is obtained as $I = 2 \cdot 4 \cdot R_s \cdot \mathrm{GMI} \approx 333.3 \cdot \mathrm{GMI}\ \mathrm{Gbit/s}$ and is reported on the right vertical axis. The performance obtained with the PAS scheme described above (solid lines) is compared with that obtained with the MB distribution (dashed lines), indicating the performance of a DM with $N \to \infty$. For computational reasons, it was not possible to consider PAS with block length longer than $N = 512$. However, we further considered the case of PAS with blocks of $512$ amplitudes followed by an $N$-symbol interleaver which randomly rearranges the amplitudes over $N/512$ adjacent blocks (shown with dotted lines extending the solid lines with same color). In this manner, the correlation is spread over $N$ adjacent symbols, as if using a longer block length, though the performance is not exactly the same—in fact, the (very small, <0.01 bits/amplitude[19]) rate loss of ESS with a block length of 512 remains also for larger $N$. The BPS was previously optimized for both cases, using $N_{\mathrm{BPS}} = 24$ for EDC and $N_{\mathrm{BPS}} = 16$ for DBP. When no carrier phase recovery is employed (only MPR), the GMI of the PAS system increases until about $N = 256$, and then slowly decreases, approaching the MB curve for $N > 4096$. The difference between the maximum of each PAS curve and the corresponding MB curve is the nonlinear shaping gain provided by PAS with optimized block length. The nonlinear shaping gain is approximately $0.085$ bits/symbol when DBP is not employed (only EDC) and $0.066$ bits/symbol when ideal DBP is employed.

Instead, when employing BPS, no nonlinear shaping gain is observed. For short block lengths $N$, the GMI is approximately the same as without BPS, but then increases monotonically and approaches the MB curve for $N \approx 128$. In fact, in this case, nonlinear interference is already mitigated by BPS, so that the optimal performance is already achieved by the MB curve and no additional gain is obtained by PAS with optimized block length. In general, Fig. 2 shows that (i) PAS and BPS mitigate the same inter-channel nonlinear interference (the former reducing its generation at the transmitter, the latter compensating it at the receiver); (ii) the combination of PAS and BPS does not offer any additional gain; (iii) when BPS is employed, the system performance does not significantly worsen when employing very long block lengths (or interleavers); (iv) with DBP (which, on the other hand, addresses intra-channel nonlinearity), the overall performance is improved and the nonlinearity mitigation induced by either BPS or short block length PAS is maintained.

All in all, Fig. 2 conveys that the nonlinear shaping gain induced by PAS can be equivalently obtained with a simple BPS algorithm. However, when BPS is used for carrier phase recovery, it may not be optimized also for inter-channel interference. For example, when the SNR is very low, the BPS should use a much larger $N_{\text{BPS}}$ to properly average out the noise and remove inter-channel interference. This scenario is considered in Fig. 3, obtained in the same scenario as Fig. 2 but considering $27$ spans of fiber, rather than only $15$. In this case, $N_{\text{BPS}} = 92$ was used. When ideal DBP is considered, the behavior of the system is similar to Fig. 2: a nonlinear shaping gain ($0.16$ bits/symbol in this case) is obtained only when BPS is not employed. Conversely, without DBP, the figure shows that both the cases with and without BPS obtain some nonlinear shaping gain ($0.1$ bits/symbol and $0.15$ bits/symbol, respectively) when the PAS block length is optimized. Furthermore, Fig. 3 shows that both PAS and MB curves obtained with BPS lie below the ones without BPS, indicating that BPS can not properly average out noise in this case, and a longer (but impractical) $N_{\text{BPS}}$ may be required.

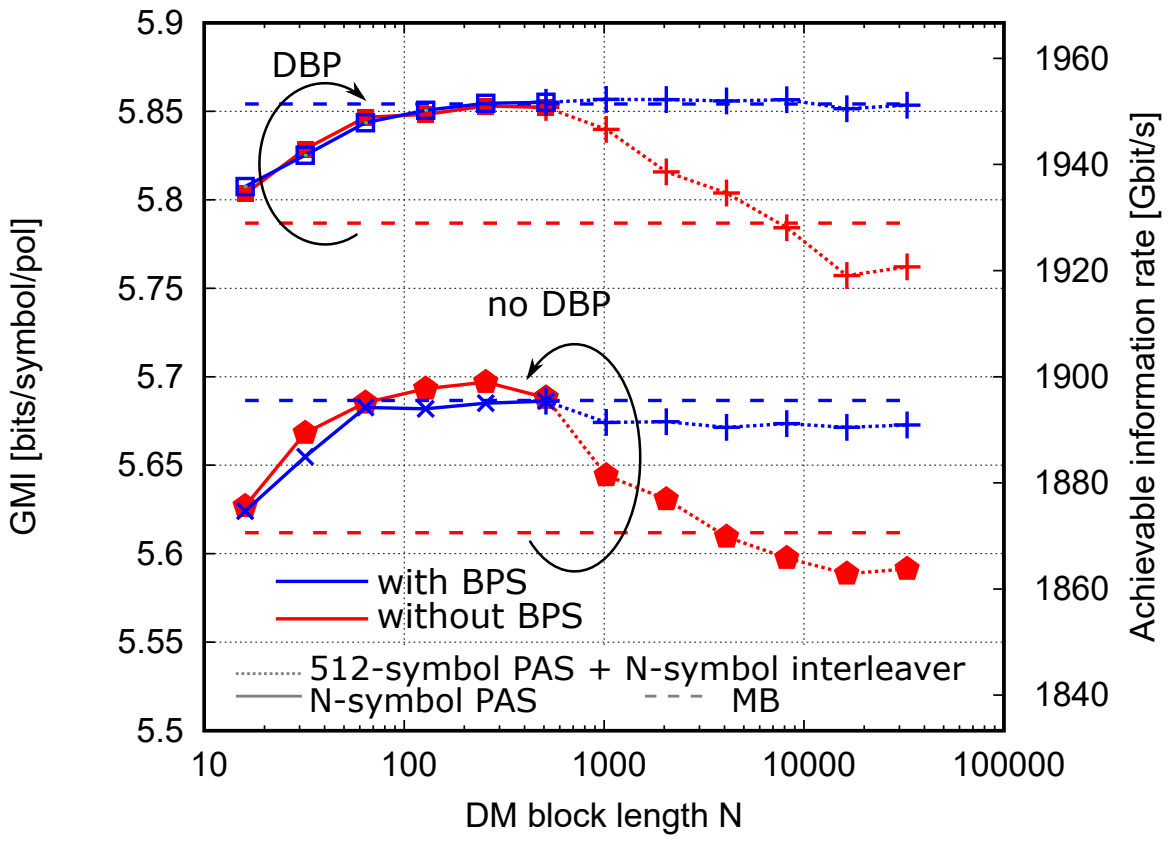

**Fig. 2:** Maximum average GMI (left axis) and total achievable information rate over the SCOI (right axis) versus DM block length $N$ for $15$ spans. Solid lines denote the performance versus $N$, dotted lines mimic the behaviour for longer $N$, and dashed lines denote the performance with i.i.d. MB symbols.

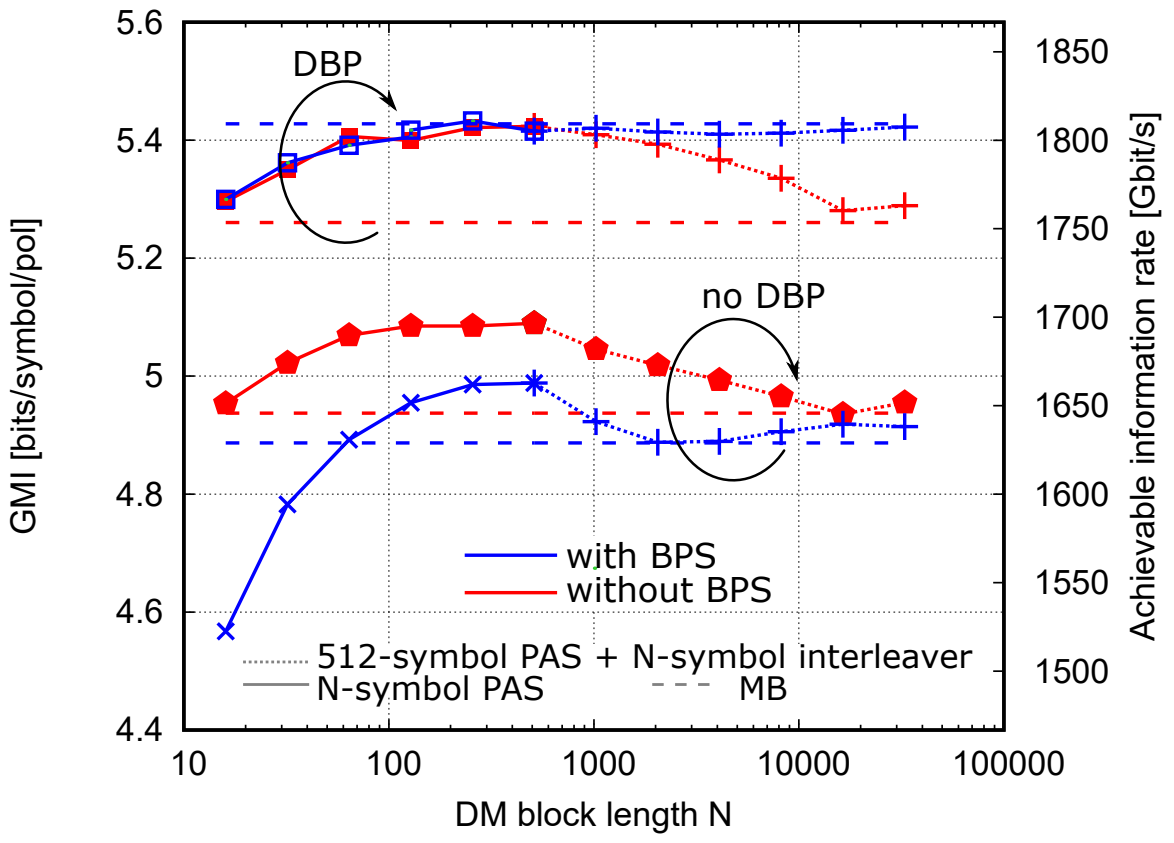

**Fig. 3:** Same as Fig. 2 but for $27$ spans.

## Conclusions

The interaction between PAS and carrier phase recovery algorithms in the nonlinear regime has been investigated for the first time. It was shown that, for high SNR, an optimized-length PAS and the common BPS algorithm typically used for carrier phase recovery can mitigate the same amount of inter-channel nonlinear interference. In this case, no additional gain is obtained by their combination. On the other hand, for low SNR, the BPS capability to mitigate a rapidly varying nonlinear phase noise is reduced, and its combination with an optimized-length PAS can provide an additional nonlinear shaping gain. Therefore, we generally recommend to include the carrier phase recovery algorithm when studying and optimizing the performance of PAS in the nonlinear regime.